\documentclass[aps,pre,preprint,groupedaddress]{revtex4}

	\usepackage{subfigure}
	\usepackage{epsfig}
	\usepackage{epstopdf}
\usepackage{caption}
\usepackage{graphicx}

\begin{document}


\title{Scaling Properties of Urban Facilities}


\author{Liang Wu}
\affiliation{School of Economics, Sichuan University, Wangjiang Road 29, Chengdu, Sichuan, 610065, P. R. China}


\date{\today}

\begin{abstract}
Two measurements are employed to quantitatively investigate the scaling properties of the spatial distribution of 
urban facilities, the $K$ function by number counting and the variance-mean relationship with the method of 
expanding bins. The $K$ function and the variance-mean relationship are both power functions. It means that 
the spatial distribution of urban facilities are scaling invariant. Further analysis of more data (which includes 
8 types of facilities in 37 major Chinese cities) shows that the exponents of the power function do not have systematic 
variations across facilities and cities, which suggests the possibility that the scaling rule is universal. 
A double stochastic process (DSP) model is proposed such that the two empirical results can both be embedded. 
Simulation of DSP yields better agreement with the urban data than of the correlated percolation model. 
\end{abstract}


\maketitle

\section{Introduction}
There has been an increasing interest to study cities and urban lives mainly due to our increasing ability to collect data which gives us better clues how cities are functioning. Some empirical regularities have been established about cities. Most urban properties, such as, land area, socio-economic rate, etc, vary continuously with population size and are well described mathematically as power-law scaling relations\cite{FractalCity,Bettencourt}. The size distribution of cities also fits a power function (known as Zipf's law): the number of cities with populations greater than $S$ is proportional to $1/S$\cite{CityZipf}. 
Geometrically, the complex spatial structure of cities have apparent fractal nature associated with individual cities and entire urban systems\cite{NatureUrbanGrowth}. 

Most traditional research of city geography focus on cities or city clusters. Providers of electronic maps such as Google 
and Baidu give us access to spatial structure data at the sub-scales of cities. These data records spatial coordinate
information of numerous urban facilities. 
China has been experiencing the largest urbanization process in human history. About 300 million people 
move to cities in the last 2 decades\cite{ChinaUrban}. As a response, many urban facilities have been developed 
to satisfy their needs. It is of special interest how these facilities are spatially organized during their rapid 
development. 

In this paper, we present evidence that spatial distribution of urban 
facilities, much alike that of city clusters, are statistically self-similar at all scales. Two measurements are 
employed to confirm these findings. The first measurement is the second-order statistics $K$ function by number counting\cite{DalyeBook}. 
Derivative of $K$ function gives the pair correlation and covariance functions. If the deviation of $K$ from its 
complete spatial randomness(CSR) version $\pi t^2$ is a power function $K(t)=\pi t^2 + K_0t^f$, it means that the pair 
correlation is also a power function. The second measurement is the method of expanding bins. The variance $V$ and 
average number $M$ of events in a series of expanding bins are related by a power function $V = aM^b$, which also
implies a self-similar scaling property of urban facilities\cite{MethodOfBins}. 
The two methods are closely related to each other. 
The exponents $b$ and $f$ are found to satisfy $b = 1+f/2$ in our empirical results. However, they are not 
completely redundant. Some model, e.g., the correlated percolation model (CPM) does not produce the desired variance-mean relationship even though the fitted spatial covariance function is imposed to its random field. 

One goal of this paper is to understand whether the power law of variance-mean and power law autocorrelation 
function indicate a universal scaling rule about urban facilities. By doing so, we apply the two methods to a lot 
of spatial data of urban facilities, which includes 8 facilities in 37 major Chinese cities. The power 
laws seem to hold for all combinations of facilities and cities, so does the relationship of the exponents $b=1+f/2$.
Then as the other goal of this paper, we propose a mathematical model of double stochastic process (DSP) in which 
the two empirical findings can both be embedded. The DSP model resembles the actual process of urban facilities in the 
sense that urban facilities are developing on top of the existing structure of a city while the city structure 
itself can also be modelled as a stochastic process. One possible explanation of the scaling "universality" of 
urban facilities is that they come from the same source of the scaling invariance of the environment of the city, 
which could result from the fractal nature of geographic characteristics of the city\cite{FractalGeology} or the 
fractal residential settlement\cite{Settlement}. However this picture does not rule out the possibility that the
scaling property of urban facilities is from some type of critical process arising from interactions between the
facilities and city environment and among the facilities themselves. 
The double stochastic process is a mathematical framework which models the macro statistical properties
of urban facilities and ignores the underlying interactions. 


There are two ways to generate point patterns with scaling invariance. One way is to have a lattice model in which 
the point patterns are formed according to some rules on a lattice. Diffusion limited aggregation (DLA) is such 
a model in which particles are added on at a time and move randomly until they join the cluster\cite{DLA, DLACity}. 
The model produces the desirable fractional power-law behaviour of the correlation function. One concern when 
applying DLA model to urban systems is that the treelike dendritic structures generated from DLA model does not 
resemble the actual spatial morphology\cite{CorrelatedPercolation}. Also urban systems do not have obvious central 
places as seen in DLA model. 

Another method is to assume that underlying the discrete point pattern there is a continuous random field. The 
spatial correlation properties of the point patterns can be imposed to the random field. 
Correlated percolation model(CPM) puts the idea into practice to model city growth\cite{CorrelatedPercolation}, 
which takes the development process of a city cluster as correlated rather than being added to the cluster at 
random. With slight different notations as in the paper, we summarize CPM as follows. The model generates a 
Gaussian random process $\lambda(\vec{x})$ with a long range power correlation function. By choosing a spatial 
varying occupancy probability $p(\vec{x})$, the model can control the spatial concentration of population density. 
For a realization of random field $\lambda(\vec{x})$, the discretized occupancy sequence is determined by 
$N(i,j) = \Theta(\Phi^{-1}(p(\vec{x}))-\lambda(\vec{x}))$, where $\Phi$ is the cumulative distribution function of 
the Gaussian random variable $\lambda(\vec{x})$ and $\Theta$ is the Heaviside step function.
The model is very successful in modelling both dynamics of city development and static statistical properties of 
the perimeter of the city cluster and power law distributions of urban settlements. 

It is appealing to apply CPM to urban facilities as the subscale analogy to city clusters. However, as shown in Appendix, CPM does not produce $b=1+f/2$ if the covariance structure of $K$ function is imposed to the random field $\Lambda(\vec{x})$. The scaling property of variance-mean relationship $b=1+f/2$ rely on the covariance of $\Theta(\Lambda(\vec{x})-T)$ after a threahold $T$ is set to generate point patterns. On the other hand, the two empirical scaling properties can be easily embedded in DSP model. Besides, CPM does not generate the same results when we generate the discrete point patterns and change the scale of discretization. The Heaviside function $\Theta$ is either 0 or 1. For example, when threshold $T$ is chosen, if both $\lambda(i)>T$ and $\lambda(i+1)>T$, then $N(i)=1, N(i+1)=1$. If a larger discretization scale is taken to combine $i$ and $i+1$ to one vertex, $\lambda(i')=\lambda(i)+\lambda(i+1)>T$, then $N(i')=1\neq N(i)+N(i+1)$. Thanks to the additivity of Poisson distribution, DSP gives the same results when applied to different scales of discretization. In this case, $\lambda(i')=\lambda(i)+\lambda(i+1)$, $\text{E}[N(i')]=\text{E}[N(i)]+\text{E}[N(i+1)]$. The additivity is preserved. 

In this paper, DSP and CPM are both implemented for comparison. DSP fits the variance and mean power relationship 
closer to empirical results than CPM.



\section{Methods and Data}
\subsection{Pair Correlation and $K$ function}
In the continuous limit of a point pattern, we can define a random field 
$\Lambda(\vec{x})=\lim_{|d\vec{x}|\rightarrow 0}{\frac{N(d\vec{x})}{|d\vec{x}|}}$. As the tradition, an upper case
letter is used to denote a random variable, and lower case one to denote a sample. 
In spatial point analysis\cite{DiggleBook}, the first and second order properties of point pattern are described by
its intensity function $\lim_{|d\vec{x}|\rightarrow 0}{\frac{\text{E}[N(d\vec{x})]}{|d\vec{x}|}} = \text{E}[\Lambda(\vec{x})] = \lambda(\vec{x})$, 
and second-order intensity function $\lambda_2(\vec{x},\vec{y}) \equiv \lim_{|d\vec{x}|, |d\vec{y}|\rightarrow 0}\frac{\text{E}[N(d\vec{x})N(d\vec{y})]}{|d\vec{x}||d\vec{y}|} =\text{E}[\Lambda(\vec{x})\Lambda(\vec{y})]
$. 
Covariance density function which measures the covariance of number of events at two infinitesimal regions $\vec{x}$ and $\vec{y}$ can be written as
\begin{eqnarray}
\gamma(t) = \lambda_2(\vec{x},\vec{y}) - m^2.
\end{eqnarray} 
where $t=|\vec{x}-\vec{y}|$. For a stationary isotropic point process, $\lambda(x)=m$ is a constant, covariance density function only depends on distance $t$ between two spatial locations.

One way to estimate the covariance density is through the $K$ function by number counting, which is defined as, 
\begin{eqnarray}
K(t) = m^{-1}\text{E}[N_0(t)],
\end{eqnarray}
where $N_0(t)$ is the number of further events within distance $t$ of an arbitrary event. It can be shown that 
$\gamma(t) = m^2((2\pi t)^{-1}K'(t)-1)$\cite{DiggleBook}.

For a point process with complete spatial randomness (CSR), one has $K(t)=\pi t^2$. 
If the deviation of $K(t)$ from CSR version is a power function, e.g., $K(t) = \pi t^2 + K_0 t^f$, then, 
\begin{eqnarray}
\label{gamma}
\gamma(t)=m^2fK_0/(2\pi)t^{-(2-f)}.
\end{eqnarray}

\subsection{Method of Expanding Bins}
Another way to analyze scaling properties of point pattern is the method of expanding bins. 
A set of equal-sized
non-overlapping bins are introduced to divide the urban area of a city into $m_i$ equal-sized segments, the size 
of each bin is $s_i^2 = A/m_i$, $A$ is the area of the whole city. The number of facilities $n_{i,j}$ is counted 
for each bin $j$. We assume that the distribution of urban facilities are homogeneous, thus, the sample variance 
and the average number of facilities in an area of size $s_i$ can be estimated as, 
$V_i = \text{Var}(n_{i,1:m_i}), M_i = \text{E}[n_{i,1:m_i}]$.
If $V_i$ and $M_i$ are related by a power function $V_i = aM_i^b$ 
as the size of bin to divide 
the city varies, it implies a statistically self-similar scaling of the spatial distribution\cite{MethodOfBins}. 
This method does not assume a stochastic process for the point pattern. If the point pattern is generated from 
stochastic model, e.g., from a underlying Poisson process with spatial non-homogeneous density
$\Lambda(x)$, the power law relationship of variance and average number of events implies that the Poisson density
is a spatially correlated with a power covariance density $\gamma(t)\sim t^{-c}$, and the two exponents are related 
by $b=2-c/2$ as will be shown in appendix. Since $\gamma(t)$ can be given by Eq. \ref{gamma}, then $f$ and $b$ are related by $b=1+f/2$.


\subsection{Data Source}
Through Baidu Map API, we record urban data of subscale structures which includes the spatial coordinates of 8 
urban facilities in the city area and adjacent counties and county-level cities of 37 major Chinese cities. 
The 8 facilities are: beauty salons, banks, stadiums, schools, pharmacy, convenient stores, restaurants and tea houses. 
The 37 major cities consist of 4 direct-controlled municipalities (Beijing, Shanghai, Chongqing and Tianjin), 30 Provincial capitals 
and sub-provincial cities and 3 other large cities. The spatial data is the latitude and longitude coordinates of each facility. 
The spatial data of the latitude and longitude spherical coordinates is converted to the plane coordinate data denoted by meters 
(data is all rounded to meter). Since it is hard to define the exact boundary of a city, we fix a central point of the city, and 
only consider those events fallen into a $2^{15}\times 2^{15}(m^2)$ square lattice centered around the central point. The bins size is 
chosen from $2^9$ meters to $2^{12}$ meters at interval of $2^6$ meters in the method of expanding bins. When calculating the $K(t)$ function, 
$t$ is chosen from $2^9$ meters to $2^{12}$ meters at increasing interval so that in log-log plot the distance somehow spreads uniformly. 

\section{Empirical Results}

\begin{figure}
        \centering
		\subfigure[$K$ functions]{\label{fig:a}\includegraphics[width=0.48\textwidth]{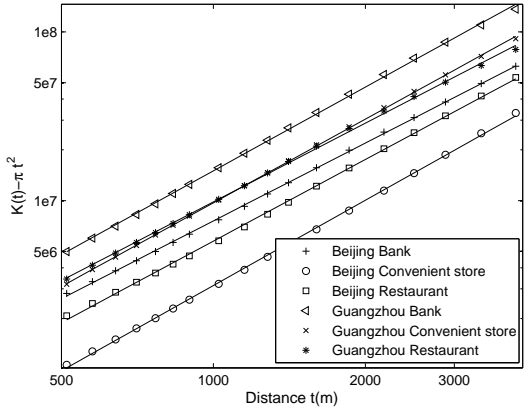}}
		\subfigure[Variance-mean relations]{\label{fig:b}\includegraphics[width=0.48\textwidth]{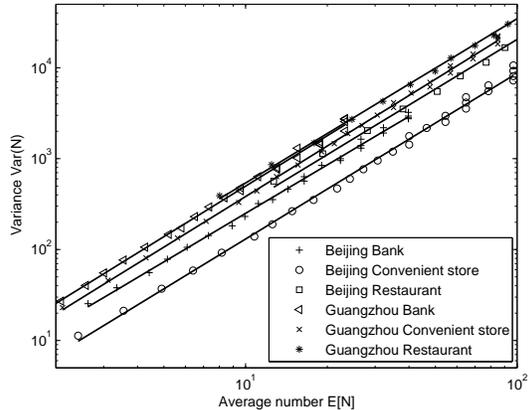}}
		\caption{Scaling properties of 3 facilities (banks, convenient stores and restaurants) in Beijing and 
		Guangzhou measured by two methods: (a)the $K$ function in a log-log plot. (b) the variance-mean relationship in a log-log plot with the 
		method of expanding bins.}\label{fig:total}
\end{figure}

As an example to show the scaling properties, we choose 3 facilities (banks, convenient stores and restaurants) 
in Beijing and Guangzhou, the largest city in the northern and southern China. There are 6 combinations out of 
3 facilities and 2 cities. As shown in Fig. \ref{fig:a}, The straight lines of $K(t)-\pi t^2$ in a log-log plot 
indicates that $K(t)-\pi t^2 = K_0 t^f$ is a power function. The exponents $f$ have very close values for 6 combinations. 
On the other hand, variance $V$ and average number of events $M$ are related by a power function $V = aM^b$ as the bin size varies in the 
method of expanding bins. The exponents $f$ and $b$ are related by $b=1+f/2$. The exact values 
of these exponents are listed in Table. \ref{tab:exponents}.

\begin{table}
\centering
\begin{tabular}{l | c | c | c}
\hline
& f & b & 1+f/2 \\
\hline
Beijing Bank                & 1.53$\pm$ 0.013 &1.77$\pm$0.04 & 1.77$\pm$ 0.006 \\
Beijing Convenient Store    & 1.68$\pm 0.018$ &1.82$\pm$0.04 & 1.84$\pm$ 0.009\\
Beijing Restaurant          & 1.61$\pm 0.021$ &1.81$\pm$0.03 & 1.80$\pm$0.010 \\
Guangzhou Bank              & 1.65 $\pm$0.020 &1.85$\pm$0.04 & 1.82 $\pm$0.010\\
Guangzhou Convenient Store  & 1.65$\pm$0.010  &1.85$\pm$0.02& 1.82 $\pm$0.005\\
Guangzhou Restaurant & 1.55$\pm$0.020&1.81 $\pm$0.03& 1.77$\pm$0.010\\ 
\hline
\end{tabular}
\caption{The exponents $f$ of the $K$ function and $b$ of the variance-mean relation, 
which are related by $b=1+f/2$. The error ranges indicate only the statistical errors from regression.}
\label{tab:exponents}
\end{table}

One goal of this paper is to understand whether the scaling invariance of urban facilities is universal. For this purpose, we apply the same 
analysis to all 8 facilities and 37 cities. The power function of both 
the $K$ function and the variance-mean relationship seem to hold for all combinations. The exponents of $b$ are 
reported in Fig. 2. In order to plot the results together, 8 cities are excluded for which there is not enough 
sample data to estimate variance-mean for at least one facility. 
We tend to believe that these exponents are 
universal in the sense that: (a) they do not have any systematic variations across 8 urban facilities regardless 
of their different nature of business or different concentrations in each city, (b) they do not show 
significant dependence on city regardless of the dramatic difference in population size, geographical 
characteristics, tradition of city planing from big cities such as Beijing to small cities such as 
Yinchuan. The error bars only indicate the statistical errors from linear regression. The variations of the 
exponents from the average value of all cities in Fig. 2 may be explained by other sources of errors. For example, the fixed choice 
of $2^{15}\times 2^{15} (m^2)$ area for each city may violate the stationery assumption as in some region of a city, e.g. a harbour city, 
there are not facilities at all.
\begin{figure}[htbp]
\centering
\subfigure[]{\label{fig:first4}\includegraphics[width=0.49\textwidth]{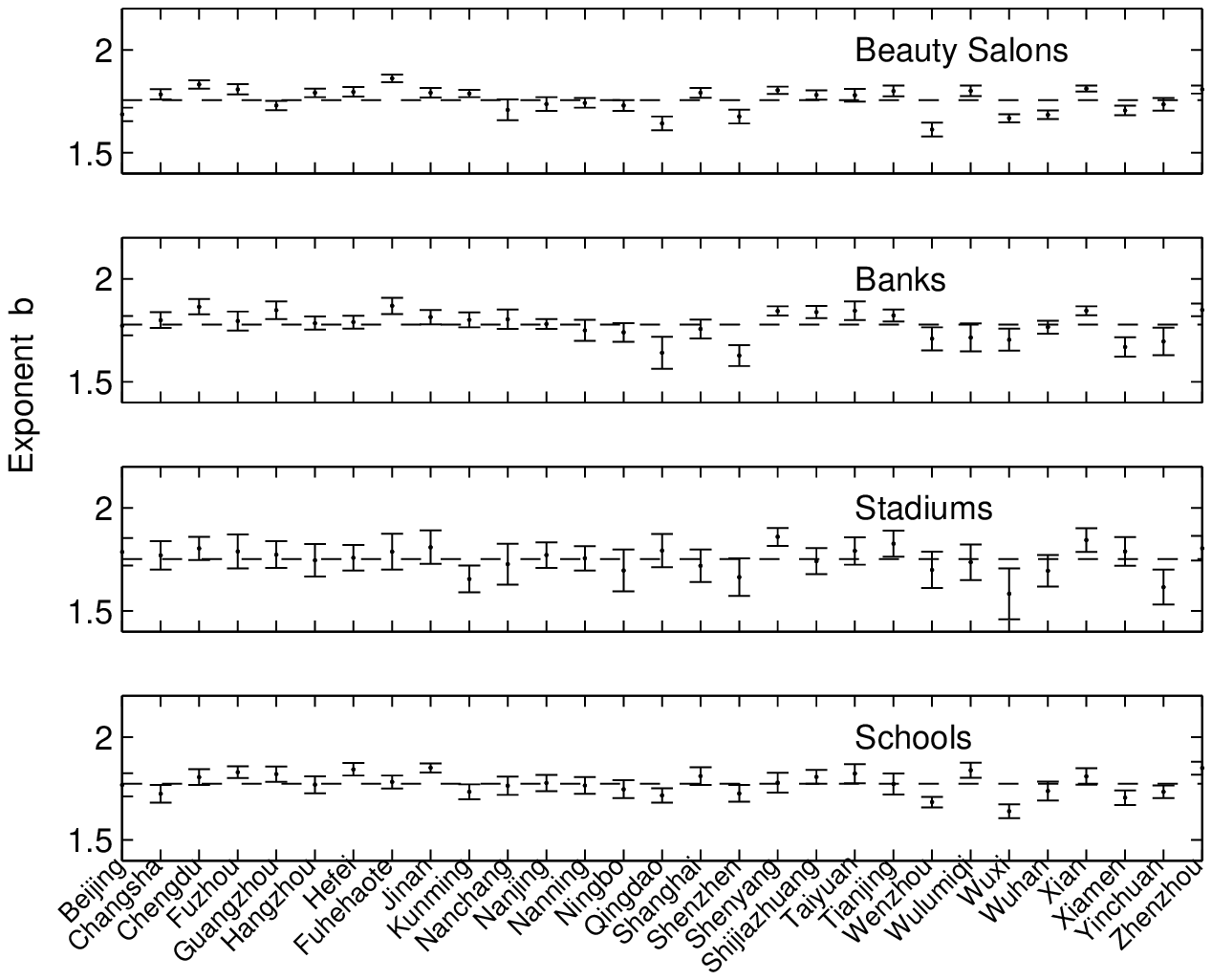}}
\subfigure[]{\label{fig:second4}\includegraphics[width=0.49\textwidth]{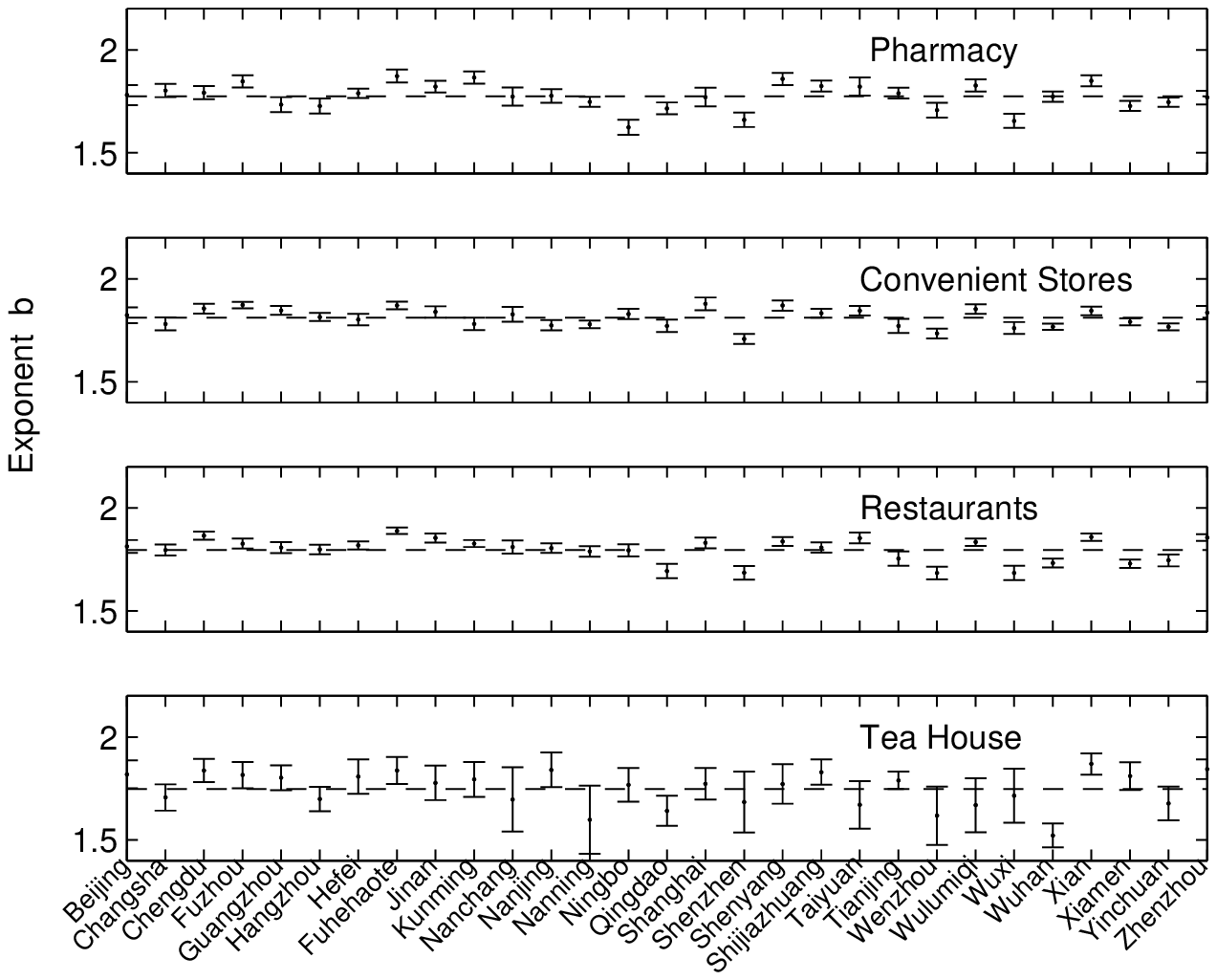}}
\caption{Exponent $b$ of the variance-mean relationship in 29 Chinese cities (8 cities are excluded for which there is not enough sample 
data for at least one facility) for (a) Beauty Salons, Banks, Stadiums, and Schools and (b) Pharmacy, Convenient Stores, Restaurants, 
and Tea Houses. The errors range indicate only the statistical errors from regression.}\label{fig:exponentB}
\end{figure}

\section{A Double Stochastic Process Model}
The other purpose of this paper is to propose a mathematical model in which the two power law rules, i.e., 
the power law of the spatial covariance and that of variance-mean relationship, can be embedded. 
It should be noted that only the first and second statistics are reflected in the power law rules. 
More information is needed, e.g., higher order statistics, in order to construct more realistic models.

CPM has been successfully applied to model the growth of city clusters. It is reasonable to visualize city 
development as addition of new units to the perimeter of an existing system. Urban facilities, on the other hand, 
is developing on top of an existing city structure. Besides the stochastic nature of the urban growth, another 
stochastic process is needed to model the randomness of the locating of urban facilities. Another reason we propose 
DSP is that it can predit the relationship of the exponents of two power law functions $b=1+f/2$.


The first layer stochastic process of DSP is a random field of density function $\Lambda(x)$, which models the 
inhomogeneous concentration of facilities in a city. The density function is a correlated random field to 
reflect the spatial heterogeneity of the city. Similar to correlated percolation model, 
the density function is long-range correlated, which takes account of the fractal structure of the city. 
Conditioned on the density, the locating of urban facilities is based on a Poisson process rather than a threshold cut off. This 
type of point process is called Cox process in the spatial point analysis. One reason to use Poisson process to generate the point 
patterns is due to the additivity of Poisson distribution. The summation of two independent Poisson distributed random variables is still 
Poisson distributed. This property is important for our model to explain the scaling invariant properties implied by the power law 
relationship between the variance and mean.


\begin{figure}[htbp]
\centering
\includegraphics[width=0.7\textwidth]{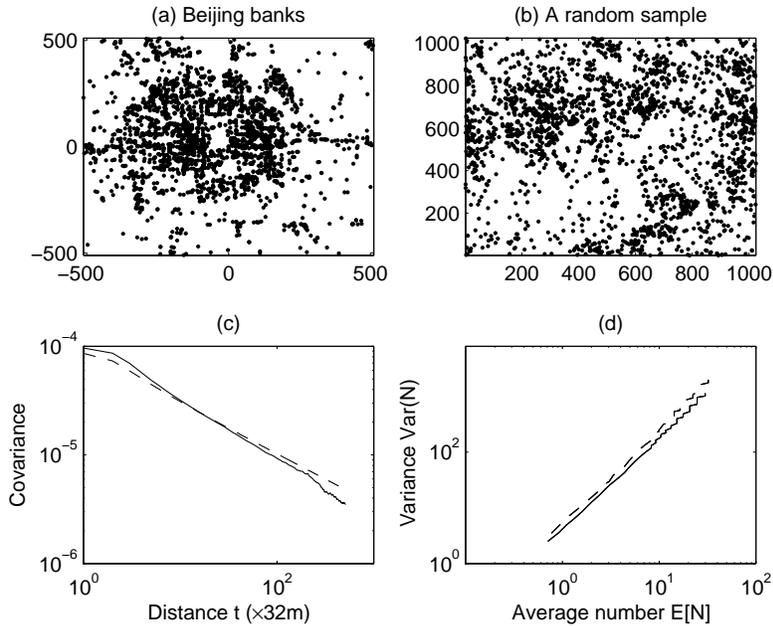}
\caption{Double stochastic process(DSP) modeling of banks in big Beijing area ($2^{15}\times2^{15}m^2$), (a) the original coordinate data of 
banks in Beijing mapped to a ($2^{10}\times2^{10}$)lattice; (b)A random sample of point pattern generated from DSP; (c)fitting of the power-law 
covariance; (d)variance-mean relationship. In both (c) and (d), dashed lines are from the real data, and solid lines are the simulation results 
averaged over 50 samples.}\label{fig:16K}
\end{figure}

\begin{figure}[htbp]
\centering
\includegraphics[width=0.7\textwidth]{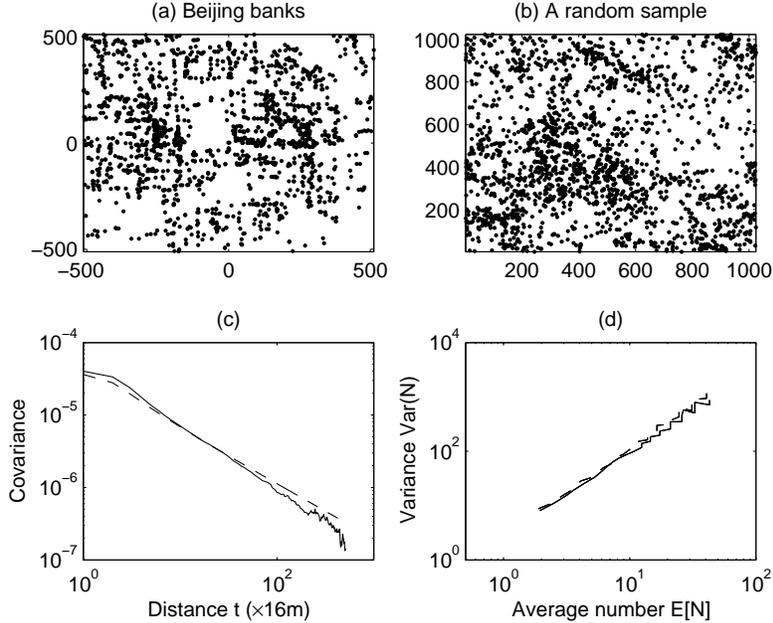}
\caption{Double stochastic process(DSP) modeling of banks in metropolitan area of Beijing ($2^{14}\times2^{14}m^2$), (a) the original coordinate 
data  of banks in Beijing mapped to a ($2^{10}\times2^{10}$)lattice; (b)A random sample of point pattern generated from DSP; (c)fitting of the 
power-law covariance; (d)variance-mean relationship. In both (c) and (d), dashed lines are results from the real data, and solid lines are those 
from simulation averaged over 50 samples.}\label{fig:8K}
\end{figure}

\begin{figure}[htbp]
\centering
\includegraphics[width=0.7\textwidth]{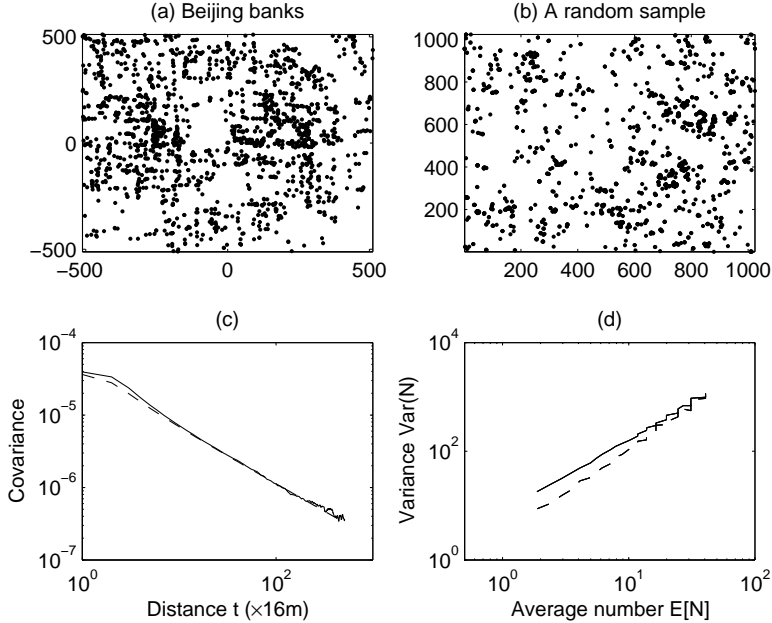}
\caption{Correlated percolation modeling(CPM) of banks in metropolitan area of Beijing ($2^{14}\times2^{14}m^2$), (a) the original coordinate 
data  of banks in Beijing mapped to a ($2^{10}\times2^{10}$)lattice; (b)A random sample of point pattern from CPM; (c)fitting of the power-law 
covariance; (d)variance-mean relationship. In both (c) and (d), dashed lines are results from the real data, and solid lines are those from 
simulation averaged over 50 samples.}\label{fig:8K_Perc}
\end{figure}

As in the appendix, we see that if the covariance of the random field $\Lambda(\vec{x})$ is a power function $\gamma(t)\propto t^{-c}$, the 
resulted point pattern, which is generated from a Poisson process conditioned on the random field as its density, is scaling invariant. 
The power law relationship between the variance and average number $V\propto M^b$ can be inherited from the power function of the covariance. 
The exponents of two power functions are related by $b = 2-c/2=1+f/2$.

A relatively flexible and tractable construction to encompass the non-negative constraint for density processes is 
log-Gaussian processes\cite{Moller}. The density function $\lambda(\vec{x})$ is drawn from a log-Gaussian random 
process $\Lambda(\vec{x}) = \text{exp}(S(\vec{x}))$. $S$ is assumed to be a 
stationery isotropic Gaussian, $S(\vec{x}) \sim \mathcal{N}(\mu, \sigma^2)$. Its spatial depedence is given by its covariance density 
$\text{Cov}(S(\vec{x}), S(\vec{x}+\vec{u})) = \sigma^2 r(|\vec{u}|)$. 
The first order and second order statistics of $S$ and $\Lambda$ fields are related by\cite{DiggleBook}, 
$m=\text{E}[\Lambda(\vec{x})] = \text{exp}(\mu+\sigma^2/2)$, and $\gamma(|\vec{u}|)=\text{Cov}(\Lambda(\vec{x}), 
\Lambda(\vec{x}+\vec{u}))=\text{exp}(2\mu + \sigma^2)[\text{exp}(\sigma^2r(\vec{u}))-1]$. Once we know the covariance $\gamma(t)$ of the 
target field $\Lambda(\vec{x})$, we can calculate the covarince of the Gaussian random field $S(\vec{x})$
field $\sigma^2r(t) = \log((2\pi t)^{-1}K'(t))=\log(\frac{{\gamma(t)}}{m^2}+1)$.

In summary, the spatial point pattern is generated in two steps: (1)a Gaussian random field $S(\vec{x})$ is sampled with desired covariance 
structure. $\Lambda(\vec{x})=\exp(S(\vec{x}))$ is thus obtained which is the density function of the point pattern;
(2)a point pattern is generated for each vertex in the lattice independently. An interger $N(\vec{x})$ is assigned to each vertex $\vec{x}$ 
following a Poisson distribution $\text{Pois}(\lambda(\vec{x}))$.

\section{Simulations}
We take banks in Beijing as an example. If the precision of the planar coordinate data is set to be 1 meter, the original data of the big 
Beijing area is on a $2^{15}\times 2^{15}$ lattice, too big for a PC to simulate. The data is mapped to a $2^{10}\times 2^{10}$ lattice 
by taking a transformation of the coordinate $x$ of each point as $x'=[x/2^5]$, rounded to its nearest integer. Its $K$ function is 
estimated as $K(t)=\pi t^2+K_0t^f$. Knowing the total number of sample points $n$ in the system, one can estimate $m$ as $m=n/2^{20}$. 

The sample point pattern is generated from two steps as described above. First, an isotropic and stationary Gaussian random field $S_{ij}$ 
is generated with the covariance function given as,
\begin{eqnarray}
\sigma^2r(t) = \log(\frac{K'(t)}{2\pi t})=\log(1+\frac{K_0f}{2\pi}r^{f-2})\approx\log(1+\frac{K_0f}{2\pi}(1+r^2)^{(f-2)/2})
\end{eqnarray}
Since $f<2$, $r^{f-2}$ diverges at $0$. We take an approximation for the correlation function which asymptotically has the same power-law behaviour.  A Gaussian random field with a given covariance can be generated from the Fourier filtering method \cite{LongRangeGenerator, CorrelatedPercolation}.

In order to have a $\lambda_{ij}$ field with desired expected value $m$. We set the expected value for Gaussian field as $\mu = \log(m)-\sigma^2r(0)/2$. The next step is to sample the point pattern. Number of events for each vertex is sampled independently as $N_{ij}\sim\text{Pois}(\lambda_{ij})$. 

The results are reported in Fig. \ref{fig:16K} and Fig. \ref{fig:8K}. In Fig. \ref{fig:16K}, we use data of banks in the big Beijing area. Fig. \ref{fig:16K}(a) is the spatial distribution of banks in the big Beijing area. Fig. \ref{fig:16K}(b) is a random sample from the log-normal double stochastic model by setting $K(t)=\pi t^2+K_0t^f, K_0=\exp{(3.62)}, f=1.54$, which is estimated from the real data. In Fig. \ref{fig:16K}(c), we report the covariance of the real data compared with the simulation data averaged over 50 samples. The covariance of the simulation data as denoted by the solid line is estimated from the power spectral of $\lambda(\vec{x})$ by fast Fourier transform\cite{AutoCorrPower}. We see that the covariance function fits pretty well although $\lambda(\vec{x})$ is not directly sampled. Fig. \ref{fig:16K}(d), the relationship of the variance-mean is reported in a log-log plot for the real data in comparison with the simulation result averaged over 50 samples. As we can see from the Fig. \ref{fig:16K}(d), the variance for the real data denoted by dashed line is slightly bigger than that of the simulation data. This is due to the fact that the real data of banks is highly concentrated in the metropolitan area of Beijing. In Fig. \ref{fig:16K}(a), there are not too many banks in the outer perimeter of the big Beijing area. It thus creates bigger variance when averaged over the whole big Beijing area than the simulation data. While in simulation, we assume the density field is stationery. One can cope with this problem by choosing a spatial varying adjustment to the field. It is not the topic of this paper. Instead, we constrain the data to the metropolitan area of Beijing which covers $2^{14}\times 2^{14} (m^2)$ and redo the simulation. As shown in Fig. \ref{fig:8K}(d), the fitting of the variance-mean is much better.

As comparison, we generate the point pattern from the correlated percolation model which is a single stochastic process. 
Now, $\gamma(t)$ is directly fed in as a covariance function to generate a Gaussian random field $\lambda(x)$. The discrete point pattern is 
then generated by applying a threshold $T = \Phi^{-1}(1-m)$ to the Gaussian field. We set $N_{ij}=1$ if $\lambda_{i,j}>T$. The experiment 
is run on the metropolitan area of Beijing. The result is reported in Fig. \ref{fig:8K_Perc}. As we can see from  Fig. \ref{fig:8K_Perc}(b), 
the resulted point pattern is highly aggregated and form isolated clusters. By setting the threshold as $T = \Phi^{-1}(1-m)$, we have the exact 
same number of points as the real data. The occupancy probability which equals to $m=0.002$ is far below the critical concentration threshold 
of a lattice percolation system. The resulted point pattern is composed of isolated clusters. We can see from Fig. \ref{fig:8K_Perc}(d), the 
variance of the simulation data, denoted by a solid line is much bigger than that of the real data. It does not fit as well as the double 
stochastic model.

\section{Conclusion and discussion}

In this paper, we use a lot of spatial data of urban facilities in Chines major cities, which are closely related to people's everyday 
lives to investigate their scaling properties. One purpose of this paper is to understand whether the spatial distribution of urban 
facilities are scaling invariant and whether the scaling rule is universal. 
Two measurments are employed to quantitatively investigate the scaling properties of the spatial distribution of urban facilities, the $K$ 
function from the spatial analysis and variance-mean relationship from the method of expanding bins. The $K$ function and the variance-mean 
relationship are both power functions, which indicate that the spatial distribution of urban facilities are scaling invariant. Further 
analysis of 8 facilities in 37 major Chinese cities shows that the exponents of the power function do not have systematic variations with 
city or facility, which suggests the possibility that the scaling rule is universal. In addition, the exponent $f$ of the $K$ function and 
$b$ of the variance-mean are related by $b=1+f/2$. The two measurements are not completely redundant. Some model, e.g. the correlated 
percolation model(CPM) does not produce the desired variance-mean relationship although the fitted spatial covariance function is imposed 
to its random field. 

The other purpose of this paper is to propose a double stochastic process (DSP) model in which the two power law rules can be embedded. 
The DSP model assumes that there is a correlated random field underlying the spatial point pattern. CPM successfully puts the idea of random  
field to model city growth. However, the cut off by applying a threshold in CPM does not preserve the additivity of random field during the 
scale change, so that it can not be applied to different scales of discretization. The other reason is that CPM does not predict the desired relationship between the two exponents $b=1+f/2$ in the point patterns generated from the random field $\Lambda(\vec{x})$ to which the covariance function implied from the empirical $K$ function is imposed. Comparison between the two models are made with simulations. The results from DSP model agree 
better with real urban data than those from CPM.

Although the assumption of the existence of a scaling invariant random field is for mathematical convenience, it resembles the actual 
process of urban facilities in the 
sense that urban facilities are developing on top of the existing structure of a city while the city structure 
itself can also be modeled as a stochastic process. One possible explanation of the scaling "universality" of 
urban facilities is that they come from the same source of the scaling invariance of the environment of the city. However this picture does not 
rule out the possibility that the
scaling property of urban facilities is from some type of critical process arising from interactions between the
facilities and city environment and among the facilities themselves. There are a broad range of physical systems which the power law scaling 
rule is discovered. The best known examples are matters near their critical point of second-order phase transition\cite{FisherReview}. 
Other examples are long polymers\cite{Polymer}, smoke-particle aggregates\cite{LongRangeSmoke}, self-avoiding walk\cite{Polymer}. 
Our findings of the power-law correlation among urban facilities do not imply a critical process. It is likely that it can be explained by 
a simple model such as diffusion-limited aggregation\cite{DLA}. Or it is a new type that has not been studied. A wide range of investigations 
of other cities around the world should be employed to explore the generality of our empirical results and the validity of the DSP modelling.





\begin{acknowledgements}
The partial financial support from the Fundamental Research Funds for the Central Universities under grant number skyb201403, and the Start up Funds from Sichuan University under grant number yj201322 is gratefully acknowledged.


\end{acknowledgements}

\section*{Appendix: the power relationship of variance-mean as a result of the power spatial covariance in the framework of a double stochastic process}
\label{appendix}

Assume that the spatial point patterns are generated from a double stochastic process. There is a random field in 2-dimensional space which has been imposed with a given covariance structure. Based on the random field as the Poisson density, a point pattern is generated from a Poisson process. 
From the additivity of Poisson distribution, the number of events $N(A)$ in region $A$ is $\text{Poiss}(\int_{A}\lambda(\vec{x})d\vec{x})$ 
for a given sample of density function $\lambda(\vec{x})$. 
\begin{eqnarray}
\text{E}[N(A)] &=& \int_{A}\text{E}[\Lambda(\vec{x})] d\vec{x} \\
\text{Var}({N(A)}) &=& \text{E}[N(A)^2]-(\text{E}[N(A)])^2 \\
&=&\text{E}[\text{E}[N(A)^2]|\Lambda]-(\text{E}[N(A)])^2 \\
&=&\text{E}[\int_A\Lambda(\vec{x})] + \text{E}[(\int_A\Lambda(\vec{x}))^2]-(\text{E}[N(A)])^2 \\
&=& \text{E}[N(A)]+\int_{A}\int_{A}\text{Cov}(\Lambda(\vec{x}),\Lambda(\vec{x}'))d\vec{x}d\vec{x}'
\end{eqnarray}
We use the fact that for $\text{Poiss}(\lambda)$, $\text{E}[N|\lambda]=\lambda, \text{E}[N^2|\lambda] = \lambda + \lambda^2$.

If the covariance function depends only on the distance between two points $\vec{x}$ and $\vec{x'}$, i.e., $\Lambda(x)$ is isotropic and 
stationary, $\text{Cov}(\Lambda(\vec{x}),\Lambda(\vec{x}'))=\gamma(|\vec{x}-\vec{x}'|)$. 
$\text{Var}({N(A)}) = \text{E}[N(A)]+\int_{A}\int_{A}\gamma(|\vec{x}-\vec{x}'|)d\vec{x}d\vec{x}'$.
Thus the integration for a fixed $\vec{u}=\vec{x}-\vec{x}'$ is the 
volume of set $\{\vec{x}:\exists\vec{x}'\in A, \text{such that } \vec{x}-\vec{x}'=\vec{u}\}$ 
times $\gamma(|\vec{u}|)$. For simplicity, we consider a rectangular region $A=[0,L_1]\times[0,L_2]$. Then,
$-L_i \leq u_i \leq L_i, i=1,2$.  
By symmetry, non-negative vectors $\vec{u}(0\leq u_i\leq L_i, i=1,2$) account for a quarter
of all $\vec{u}$s. 
Under the constraint that $\vec{x}\in A$, $\vec{x}'\in A$ and $\vec{u}=\vec{x}-\vec{x}'$, one can have $\vec{x} \in [u_1,L_1]\times[u_2,L_2]$, 
the volume of $\vec{x}$ is therefore $(L_1-u_1)(L_2-u_2)$. The above integral can be written as, 
\begin{eqnarray}
I(A)&=&\int_{A}\int_{A}\gamma(|\vec{x}-\vec{x}'|)d\vec{x}d\vec{x}' \\
&=&4\int_0^{L_1}\int_0^{L_2}\gamma(|\vec{u}|)(L_1-u_1)(L_2-u_2)du_1du_2
\end{eqnarray}

If the spatial correlation of the random
density field $\Lambda(\vec{x})$ is scaling free, i.e., ${\gamma(|\vec{u}|)} \propto |\vec{u}|^{-c}$, the scaling property 
of $\text{Var}(N(A))$ can be inherited from the scaling property of covariance function of
$\Lambda(\vec{x})$ field. We change the scale of considered region to $A'=[0,sL_1]\times[0,sL_2]$, the corresponding integral is,
\begin{eqnarray}
I(A') &=& 4\int_{0}^{sL_1}\int_0^{sL_2}\gamma(|\vec{u}|)(sL_1-u_1)(sL_2-u_2)du_1du_2 \\
&=& 4\int_{0}^{L_1}\int_0^{L_2}\gamma(|s\vec{u}|)(sL_1-su_1)(sL_2-su_2)s^2du_1du_2 \\
&=& s^{4-c}I(A)
\end{eqnarray}
Thus, the expected value and variance of the number of sample points $N(A')$ are given by, 
\begin{eqnarray}
\text{E}[N(A')] &=& \int_{A'}\text{E}[\Lambda(\vec{x})]d\vec{x}=s^2\text{E}[N(A)] \\
\label{eq:sv}
\text{Var}({N(A')}) &=& s^2\text{E}[N(A)]+ s^{4-c}I(A) 
\end{eqnarray}

Since $d>c$, the second term of Eq.~(\ref{eq:sv}) grows faster and dominates the first term. Asymptotically, as $s\rightarrow\infty$
\begin{eqnarray}
\text{Var}({N(A')}) \propto \text{E}[N(A')]^{2-c/2}
\label{eq:scaling}
\end{eqnarray}
which gives the exponent of power law between the variance and average value $b = 2-c/2$. Since $c\geq 0$, the 
upper bound is $b \leq 2$.
The equality holds ($c=0, b=2$), when the spatial correlation is a constant.
On the other hand, the first term of Eq.~(ref{eq:sv}) equals to $\text{E}(N(A'))$, which sets the lower bound of the exponent $b \geq 1$.
In all, we have $1\leq b \le 2$.

On the other hand, if a threshold cut off $T$ is applied to $\Lambda(\vec{x})$ to generate point patterns in correlated percolation model(CPM), $N'(A)=\int_A \Theta(\Lambda(\vec{x})-T)d\vec{x}$, we have,
\begin{eqnarray}
\text{E}[N'(A)] &=& \int_{A}\text{E}[\Theta(\Lambda(\vec{x})-T)] d\vec{x} \\
\text{Var}({N'(A)}) &=& \int_{A}\int_{A}\text{Cov}(\Theta(\Lambda(\vec{x})-T),\Theta(\Lambda(\vec{x'})-T))d\vec{x}d\vec{x}'
\end{eqnarray}
If $\text{Cov}(\Lambda(\vec{x}),\Lambda(\vec{x}'))\propto |\vec{x}-\vec{x}'|^{-c}$ is imposed to $\Lambda(\vec{x})$ field, $\text{Cov}(\Theta(\Lambda(\vec{x})-T),\Theta(\Lambda(\vec{x'})-T))$ does not scale as $|\vec{x}-\vec{x}'|^{-c}$. In this case, we do not have the scaling property of Eq. \ref{eq:scaling} and therefore $b=1+f/2$ does not hold.

So far every thing is discussed in the continuous limit. For a double stochastic Poisson process, thanks to the additivity of Poisson distribution, the discretization can be applied to different scales. However, it would be a problem for CPM since the Heaviside function $\Theta$ can only produce $1$ or $0$. 


\end{document}